\documentclass[aps,prl,superscriptaddress,showpacs,floatfix,twocolumn]{revtex4}

\usepackage{amsmath}
\usepackage{graphicx}
\def\sqrtsNN{\mbox{$\sqrt{s_{NN}}$}}

\newcommand{\mean}[1]{\left\langle #1 \right\rangle}

\begin{document}

\title{Scaling properties of azimuthal anisotropy \\
       in Au+Au and Cu+Cu collisions at $\sqrt{s_{NN}}= 200$~GeV
}

\newcommand{\abilene}{Abilene Christian University, Abilene, TX 79699, U.S.}
\newcommand{\banaras}{Department of Physics, Banaras Hindu University, Varanasi 221005, India}
\newcommand{\bnl}{Brookhaven National Laboratory, Upton, NY 11973-5000, U.S.}
\newcommand{\caucr}{University of California - Riverside, Riverside, CA 92521, U.S.}
\newcommand{\charlesczech}{Charles University, Ovocn\'{y} trh 5, Praha 1, 116 36, Prague, Czech Republic}
\newcommand{\ciae}{China Institute of Atomic Energy (CIAE), Beijing, People's Republic of China}
\newcommand{\cns}{Center for Nuclear Study, Graduate School of Science, University of Tokyo, 7-3-1 Hongo, Bunkyo, Tokyo 113-0033, Japan}
\newcommand{\colorado}{University of Colorado, Boulder, CO 80309, U.S.}
\newcommand{\columbia}{Columbia University, New York, NY 10027 and Nevis Laboratories, Irvington, NY 10533, U.S.}
\newcommand{\czechtech}{Czech Technical University, Zikova 4, 166 36 Prague 6, Czech Republic}
\newcommand{\dapnia}{Dapnia, CEA Saclay, F-91191, Gif-sur-Yvette, France}
\newcommand{\debrecen}{Debrecen University, H-4010 Debrecen, Egyetem t{\'e}r 1, Hungary}
\newcommand{\elte}{ELTE, E{\"o}tv{\"o}s Lor{\'a}nd University, H - 1117 Budapest, P{\'a}zm{\'a}ny P. s. 1/A, Hungary}
\newcommand{\fit}{Florida Institute of Technology, Melbourne, FL 32901, U.S.}
\newcommand{\fsu}{Florida State University, Tallahassee, FL 32306, U.S.}
\newcommand{\gsu}{Georgia State University, Atlanta, GA 30303, U.S.}
\newcommand{\hiroshima}{Hiroshima University, Kagamiyama, Higashi-Hiroshima 739-8526, Japan}
\newcommand{\ihepprot}{IHEP Protvino, State Research Center of Russian Federation, Institute for High Energy Physics, Protvino, 142281, Russia}
\newcommand{\illuiuc}{University of Illinois at Urbana-Champaign, Urbana, IL 61801, U.S.}
\newcommand{\instpasczech}{Institute of Physics, Academy of Sciences of the Czech Republic, Na Slovance 2, 182 21 Prague 8, Czech Republic}
\newcommand{\isu}{Iowa State University, Ames, IA 50011, U.S.}
\newcommand{\jinrdubna}{Joint Institute for Nuclear Research, 141980 Dubna, Moscow Region, Russia}
\newcommand{\kaeri}{KAERI, Cyclotron Application Laboratory, Seoul, South Korea}
\newcommand{\kek}{KEK, High Energy Accelerator Research Organization, Tsukuba, Ibaraki 305-0801, Japan}
\newcommand{\kfki}{KFKI Research Institute for Particle and Nuclear Physics of the Hungarian Academy of Sciences (MTA KFKI RMKI), H-1525 Budapest 114, POBox 49, Budapest, Hungary}
\newcommand{\korea}{Korea University, Seoul, 136-701, Korea}
\newcommand{\kurchatov}{Russian Research Center ``Kurchatov Institute", Moscow, Russia}
\newcommand{\kyoto}{Kyoto University, Kyoto 606-8502, Japan}
\newcommand{\labllr}{Laboratoire Leprince-Ringuet, Ecole Polytechnique, CNRS-IN2P3, Route de Saclay, F-91128, Palaiseau, France}
\newcommand{\lawllnl}{Lawrence Livermore National Laboratory, Livermore, CA 94550, U.S.}
\newcommand{\losalamos}{Los Alamos National Laboratory, Los Alamos, NM 87545, U.S.}
\newcommand{\lpc}{LPC, Universit{\'e} Blaise Pascal, CNRS-IN2P3, Clermont-Fd, 63177 Aubiere Cedex, France}
\newcommand{\lund}{Department of Physics, Lund University, Box 118, SE-221 00 Lund, Sweden}
\newcommand{\muenster}{Institut f\"ur Kernphysik, University of Muenster, D-48149 Muenster, Germany}
\newcommand{\myongji}{Myongji University, Yongin, Kyonggido 449-728, Korea}
\newcommand{\nagasaki}{Nagasaki Institute of Applied Science, Nagasaki-shi, Nagasaki 851-0193, Japan}
\newcommand{\newmex}{University of New Mexico, Albuquerque, NM 87131, U.S. }
\newcommand{\nmsu}{New Mexico State University, Las Cruces, NM 88003, U.S.}
\newcommand{\ornl}{Oak Ridge National Laboratory, Oak Ridge, TN 37831, U.S.}
\newcommand{\orsay}{IPN-Orsay, Universite Paris Sud, CNRS-IN2P3, BP1, F-91406, Orsay, France}
\newcommand{\peking}{Peking University, Beijing, People's Republic of China}
\newcommand{\pnpi}{PNPI, Petersburg Nuclear Physics Institute, Gatchina, Leningrad region, 188300, Russia}
\newcommand{\riken}{RIKEN, The Institute of Physical and Chemical Research, Wako, Saitama 351-0198, Japan}
\newcommand{\rikjrbrc}{RIKEN BNL Research Center, Brookhaven National Laboratory, Upton, NY 11973-5000, U.S.}
\newcommand{\rikkyo}{Physics Department, Rikkyo University, 3-34-1 Nishi-Ikebukuro, Toshima, Tokyo 171-8501, Japan}
\newcommand{\saispbstu}{Saint Petersburg State Polytechnic University, St. Petersburg, Russia}
\newcommand{\saopaulo}{Universidade de S{\~a}o Paulo, Instituto de F\'{\i}sica, Caixa Postal 66318, S{\~a}o Paulo CEP05315-970, Brazil}
\newcommand{\seoulnat}{System Electronics Laboratory, Seoul National University, Seoul, South Korea}
\newcommand{\stonybrkc}{Chemistry Department, Stony Brook University, Stony Brook, SUNY, NY 11794-3400, U.S.}
\newcommand{\stonycrkp}{Department of Physics and Astronomy, Stony Brook University, SUNY, Stony Brook, NY 11794, U.S.}
\newcommand{\subatech}{SUBATECH (Ecole des Mines de Nantes, CNRS-IN2P3, Universit{\'e} de Nantes) BP 20722 - 44307, Nantes, France}
\newcommand{\tenn}{University of Tennessee, Knoxville, TN 37996, U.S.}
\newcommand{\titech}{Department of Physics, Tokyo Institute of Technology, Oh-okayama, Meguro, Tokyo 152-8551, Japan}
\newcommand{\tsukuba}{Institute of Physics, University of Tsukuba, Tsukuba, Ibaraki 305, Japan}
\newcommand{\vandy}{Vanderbilt University, Nashville, TN 37235, U.S.}
\newcommand{\waseda}{Waseda University, Advanced Research Institute for Science and Engineering, 17 Kikui-cho, Shinjuku-ku, Tokyo 162-0044, Japan}
\newcommand{\weizmann}{Weizmann Institute, Rehovot 76100, Israel}
\newcommand{\yonsei}{Yonsei University, IPAP, Seoul 120-749, Korea}
\affiliation{\abilene}
\affiliation{\banaras}
\affiliation{\bnl}
\affiliation{\caucr}
\affiliation{\charlesczech}
\affiliation{\ciae}
\affiliation{\cns}
\affiliation{\colorado}
\affiliation{\columbia}
\affiliation{\czechtech}
\affiliation{\dapnia}
\affiliation{\debrecen}
\affiliation{\elte}
\affiliation{\fit}
\affiliation{\fsu}
\affiliation{\gsu}
\affiliation{\hiroshima}
\affiliation{\ihepprot}
\affiliation{\illuiuc}
\affiliation{\instpasczech}
\affiliation{\isu}
\affiliation{\jinrdubna}
\affiliation{\kaeri}
\affiliation{\kek}
\affiliation{\kfki}
\affiliation{\korea}
\affiliation{\kurchatov}
\affiliation{\kyoto}
\affiliation{\labllr}
\affiliation{\lawllnl}
\affiliation{\losalamos}
\affiliation{\lpc}
\affiliation{\lund}
\affiliation{\muenster}
\affiliation{\myongji}
\affiliation{\nagasaki}
\affiliation{\newmex}
\affiliation{\nmsu}
\affiliation{\ornl}
\affiliation{\orsay}
\affiliation{\peking}
\affiliation{\pnpi}
\affiliation{\riken}
\affiliation{\rikjrbrc}
\affiliation{\rikkyo}
\affiliation{\saispbstu}
\affiliation{\saopaulo}
\affiliation{\seoulnat}
\affiliation{\stonybrkc}
\affiliation{\stonycrkp}
\affiliation{\subatech}
\affiliation{\tenn}
\affiliation{\titech}
\affiliation{\tsukuba}
\affiliation{\vandy}
\affiliation{\waseda}
\affiliation{\weizmann}
\affiliation{\yonsei}
\author{A.~Adare}	\affiliation{\colorado}
\author{S.~Afanasiev}	\affiliation{\jinrdubna}
\author{C.~Aidala}	\affiliation{\columbia}
\author{N.N.~Ajitanand}	\affiliation{\stonybrkc}
\author{Y.~Akiba}	\affiliation{\riken} \affiliation{\rikjrbrc}
\author{H.~Al-Bataineh}	\affiliation{\nmsu}
\author{J.~Alexander}	\affiliation{\stonybrkc}
\author{A.~Al-Jamel}	\affiliation{\nmsu}
\author{K.~Aoki}	\affiliation{\kyoto} \affiliation{\riken}
\author{L.~Aphecetche}	\affiliation{\subatech}
% ...
\collaboration{PHENIX Collaboration} \noaffiliation

\date{\today}


\begin{abstract}

Detailed differential measurements of the elliptic flow for particles
produced in Au+Au and Cu+Cu collisions at $\sqrt{s_{NN}} = 200$~GeV 
are presented. Predictions from perfect fluid hydrodynamics for the 
scaling of the elliptic flow coefficient $v_2$ with eccentricity, system 
size and transverse energy are tested and validated. For transverse kinetic 
energies $KE_T \equiv m_T - m$ up to $\sim 1$~GeV, scaling compatible with 
the hydrodynamic expansion of a thermalized fluid is observed for all 
produced particles. For large values of $KE_T$, the mesons and baryons 
scale separately. A universal scaling for the flow of both mesons and 
baryons is observed for the full transverse kinetic energy range of the 
data when quark number scaling is employed. In both cases the scaling is 
more pronounced in terms of $KE_T$ rather than transverse momentum. 
\end{abstract}



\maketitle



Quantum Chromodynamics calculations  
performed on the lattice (LQCD) indicate a transition from a low-temperature phase of 
nuclear matter, dominated by hadrons, into a high-temperature 
plasma phase of quarks and gluons (QGP)~\cite{Karsch:2001vs}. For matter 
with zero net baryon density, this phase transition has been predicted to occur 
at an energy density of  $\sim 1$~GeV/fm$^3$ or for a critical 
temperature $T_c \sim 170$~MeV \cite{Karsch:2000ps}.
Recent estimates from transverse energy ($E_T$) measurements 
at the relativistic heavy ion collider (RHIC) have indicated energy 
densities of at least $5.4$~GeV/fm$^3$ in central Au+Au 
collisions~\cite{Adcox:2004mh}. Thus, an important prerequisite 
for QGP production is readily fulfilled at RHIC. 
Indeed, there is much evidence that thermalized nuclear matter 
has been created at unprecedented energy densities in heavy ion collisions at 
RHIC~\cite{Adcox:2004mh,Adams:2005dq,Back:2004je,Arsene:2004fa,
Gyulassy:2004zy,Muller:2004kk,Shuryak:2004cy,Heinz:2001xi}. 

	Hydrodynamics provides a link between the fundamental 
properties of this matter (its equation of state (EOS) and transport coefficients) 
and the flow patterns evidenced in the measured hadron spectra and azimuthal 
anisotropy~\cite{Teaney:2001av,Huovinen:2001cy,Hirano:2004rs,Csorgo:1995bi,Csanad:2005gv}. 
Experimentally, such a momentum anisotropy is commonly characterized at mid-rapidity, 
by the even order Fourier coefficients \cite{Ollitrault:1992bk,Poskanzer:1998yz}, 
\begin{equation}
 v_n = \mean{e^{in(\phi_p - \Phi_{RP})}}, {\text{  }} n=2,4,.., 
\label{eq1}
\end{equation} 
where $\phi_{p}$ represents the azimuthal emission angle of a particle, 
$\Phi_{RP}$ is the azimuth of the reaction plane and the brackets denote 
statistical averaging over particles and events.  

	At low transverse momentum ($p_T \alt 2.0$ GeV/$c$) the magnitude and trends of 
elliptic flow, measured by the second Fourier coefficient $v_2$,
is found to be under-predicted by a hadronic cascade model~\cite{Bleicher:2000sx}. 
By contrast, a broad selection of the data showed good quantitative agreement 
with perfect fluid (very low ratio of viscosity to entropy)
hydrodynamics \cite{Shuryak:2004cy,Heinz:2001xi,Huovinen:2001cy,Csanad:2005gv} and 
a transport model calculation which incorporates extremely large opacities~\cite{Molnar:2001ux}.
For higher $p_T$, quark coalescence from a thermalized state of flowing 
partonic matter~\cite{Fries:2003kq} has been found to be consistent with 
the data~\cite{Adler:2003kt,Adams:2003am}. Together, these results 
provide evidence for the production of a strongly interacting QGP whose subsequent 
evolution is similar to that of a ``perfect" 
fluid \cite{Shuryak:2004cy,Gyulassy:2004zy,Muller:2004kk,Heinz:2001xi}.

Systematic theoretical and experimental studies of the
influence of model parameters are now required to gain more
quantitative insight on the transport coefficients and the
EOS for this strongly interacting matter. The range of
validity of perfect fluid hydrodynamics is affected by the
degree of thermalization~\cite{Bhalerao:2005mm} and the onset of dissipative
effects~\cite{Hirano:2005wx,Bhalerao:2005mm,Hirano:2005xf}. These questions can be addressed 
by investigating several scaling predictions of perfect fluid 
hydrodynamics \cite{Borghini:2005kd,Bhalerao:2005mm,Lacey:2005qq,Csanad:2005gv,Fries:2003vb}.


	In the hydrodynamic model, elliptic flow can result from 
pressure gradients due to the initial spatial asymmetry 
or eccentricity $\epsilon= (\langle y^2- x^2\rangle)/(\langle y^2+ x^2\rangle)$, 
of the high energy density matter in the collision zone. The initial 
entropy density $S(x,y)$, can be used to perform an average over the 
$x$ and $y$ coordinates of the matter in the plane perpendicular to the 
collision axis. Here, $x$ points along the impact vector and $y$ is orthogonal to $x$.
For a system of transverse size $\bar R$ ($1/\bar R=\sqrt{1/\langle x^2\rangle+1/\langle y^2\rangle}$), 
this flow develops over a time scale $\sim \bar R/\left\langle c_s\right\rangle$
for matter with an average speed of sound $c_s$. Thus, the  
initial energy density controls how much flow develops globally, while   
the detailed development of the flow patterns are largely controlled by 
$\epsilon$ and $c_s$. 

	An important prediction of perfect fluid hydrodynamics is that the relatively ``complicated" 
dependence of azimuthal anisotropy on centrality, transverse momentum, rapidity, 
particle type, higher harmonics, etc can be scaled to a single function 
\cite{Csanad:2005gv,Csanad:2006sp}.
Immediate consequences of this 
\cite{Csanad:2005gv,Csanad:2006sp,Borghini:2005kd,Bhalerao:2005mm}
are that: 
(i) $v_2$ scaling should hold for a broad range of impact parameters for which the 
eccentricity varies, i.e. $v_2(p_T)/\epsilon$ should be independent of centrality; 
(ii) $v_2(p_T)$ should be independent of colliding system size for a given 
eccentricity; and
(iii) for different particle species, $v_2(KE_T)$ at mid-rapidity should scale 
with the transverse kinetic energy $KE_T = m_T-m$, where $m_T$ is the transverse 
mass of the particle. 

	We use high statistics $v_2$ data 
to test these scaling predictions and explore constraints for 
the range of validity of perfect fluid hydrodynamics.  
The measurements were made at $\sqrt{s_{NN}} = 200$ GeV with the PHENIX 
detector~\cite{Adcox:2003zm} at RHIC. 
Approximately $6.5 \times 10^8$ Au+Au and 
and $8.0 \times 10^7$ Cu+Cu 
minimum-bias collisions were analyzed
from the 2004 and 2005 running periods, respectively.
The collision vertex $z$, along the beam direction 
was constrained to be within $|z| <$~30~cm.
The event centrality for Au+Au collisions was determined 
via cuts in the space of Beam-Beam Counter 
(BBC) versus Zero Degree Calorimeter analog response~\cite{Adcox:2003nr}. 
For Cu+Cu only the amplitude of the BBC analog response was used. 
Charged hadrons were detected in the two central arms ($|\eta|\leq 0.35$). 
Track reconstruction was accomplished using the drift chambers and two 
layers of multi-wire proportional chambers with pad readout (PC1 and PC3) 
located at radii of 2 m, 2.5 m and 5 m, respectively \cite{Adcox:2003zm}. 


The time-of-flight (TOF) detector positioned at a radial distance of 5.06 m,  
was used to identify pions ($\pi^{\pm}$), kaons ($K^{\pm}$) 
and (anti)protons $(\overline{p})p$. The BBCs and TOF-scintillators 
provided the global start and stop signals.
These measurements were used in conjunction with 
the measured momentum and flight-path length to generate a 
mass-squared distribution \cite{Adler:2003cb}. 
A momentum dependent $\pm 2\sigma$ cut about each 
peak in this distribution was used to identify $\pi^{\pm}$, $K^{\pm}$ 
and $(\overline{p})p$ in the range $0.2<p_T<2.5$ GeV/$c$, $0.2<p_T<2.5$ GeV/$c$ 
and $0.5<p_T<4.5$ GeV/$c$, respectively. A track confirmation hit within 
a $2.5 \sigma$ matching window in PC3/TOF served to eliminate most albedo, 
conversions, and resonance decays.    

	The differential elliptic flow measurements for charged hadrons and identified particles were 
obtained with the reaction plane method. This technique correlates the azimuthal 
angles of charged tracks 
with the azimuth of the event plane $\Phi_{2}$, determined via hits in the two BBCs
positioned symmetrically along the beam line,  
covering the pseudo-rapidity range $3 < \left| \eta \right| < 3.9 $ \cite{Adler:2003kt}. 
A large $\eta$ gap between the central arms 
and the particles used for reaction plane determination reduces 
the influence of possible non-flow contributions, especially those from jets.
Values of $v_2$ were calculated via the expression
\begin{equation}
v_2 = {\frac{\left\langle \cos(2(\phi_p -\Phi_2))\right\rangle}
{\left\langle\cos(2(\Phi_2-\Phi_{RP}))\right\rangle}},
\end{equation}
where the denominator represents a resolution factor that corrects for
the difference between the estimated $\Phi_{2}$ and the true azimuth $\Phi_{RP}$ of the 
reaction plane \cite{Adler:2003kt,Adler:2001nb}. 
The estimated resolution factor of the combined reaction 
plane from both BBCs~\cite{Adler:2003kt} 
has an average of 0.33 (0.16) 
over centrality with a maximum of about 0.42 (0.19) for Au+Au (Cu+Cu).
The estimated correction factor for the $v_2$ measurements
(i.e. the inverse of the resolution factor) ranges from 2.4 (5.5) 
to 5.0 (13). Relative systematic errors for these $v_2$ values are estimated to 
be $\sim 5$\% and $\sim 10$\% for Au+Au and Cu+Cu, respectively.

\begin{figure}[tb]
\includegraphics[width=1.0\linewidth]{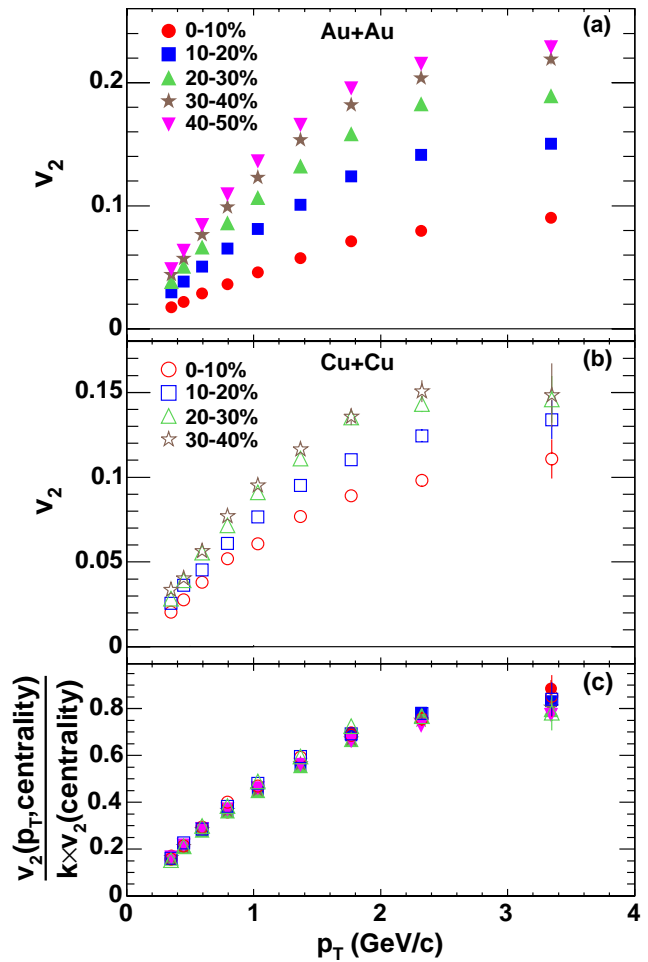}
\caption[]{\label{fig:ecc_scaling}
$v_2$ vs. $p_T$ for charged hadrons obtained in (a) Au+Au and
(b) Cu+Cu collisions for the centralities indicated.  (c) 
$v_2(centrality, p_T)$ divided by $k$=3.1 (see text) times the 
$p_T$-integrated value $v_2(centrality)$ for Au+Au and Cu+Cu. 
}
\end{figure}

Figure \ref{fig:ecc_scaling} shows the differential $v_2(p_T)$ 
for charged hadrons obtained in Au+Au and Cu+Cu collisions.
The $v_2(p_T)$ results 
exhibit the 
familiar increase as collisions become more peripheral and the $p_T$ 
increase \cite{Adcox:2004mh,Adams:2005dq,Back:2004je}. 
We test these data for eccentricity scaling by 
dividing the differential values shown in Fig. \ref{fig:ecc_scaling} 
by the $v_2$ integrated over the $p_T$ range 0.3-2.5 GeV/$c$
for each of the indicated centrality selections. 
The hydrodynamic model predicts that this ratio is constant with 
centrality and independent of colliding system because $\epsilon$ is proportional 
to the $p_T$-integrated $v_2$ values (i.e. $\epsilon = k \times v_2$). 
The latter proportionality has been observed for Au+Au 
collisions \cite{Ackermann:2000tr,Adcox:2002ms}.
A Glauber model estimate of $\epsilon$ \cite{Adcox:2002ms} gives $k = 3.1 \pm 0.2$ for 
the cuts employed in this analysis.
%
This method of scaling leads to a scale invariant variable and cancels  
the systematic errors associated with estimates of the reaction plane  
resolution and the eccentricity.

	The resulting scaled $v_2$ values for Cu+Cu and Au+Au collisions, are shown 
in Fig.~\ref{fig:ecc_scaling}(c). To facilitate later comparisons with the 
model calculations of Ref.~\cite{Bhalerao:2005mm}, they are divided by $k=3.1$.
These scaled values are clearly independent of the colliding system 
size and show essentially perfect scaling for the full range of 
centralities (or $\epsilon$) presented. 
The $v_2$ are also in accord with the scale invariance of perfect fluid 
hydrodynamics \cite{Lacey:2005qq,Bhalerao:2005mm}, which suggests that 
rapid local thermalization \cite{Shuryak:2004cy,Heinz:2001xi} is 
achieved.
 
%
The magnitude of $v_2/\epsilon$ depends on the sound speed $c_s$ \cite{Bhalerao:2005mm}.  
As a reasonable first approximation we compare our measured $v_2/\epsilon$ at an 
integrated $\left\langle p_T \right\rangle$~0.45 GeV/c and the results of Fig. 2 
of \cite{Bhalerao:2005mm}.  This results in a speed of sound $c_s \sim 0.35 \pm 0.05$.  
Note that the calculations are done at fixed b=8 fm and a constant speed 
of sound.  Thus, since we expect the speed of sound to vary as a function 
of time, one might view this $c_s$ value as the approximate average value over 
the time period 2 $\bar R/c_s$, the time over which the flow develops.
This value suggests an effective EOS, which is softer than that for the 
high temperature QGP \cite{Karsch:2006sm} but does not reflect a strong 
first order phase transition in which $c_s = 0$ during an extended 
hadronization period.

%
\begin{figure}[tb]
\includegraphics[width=1.0\linewidth]{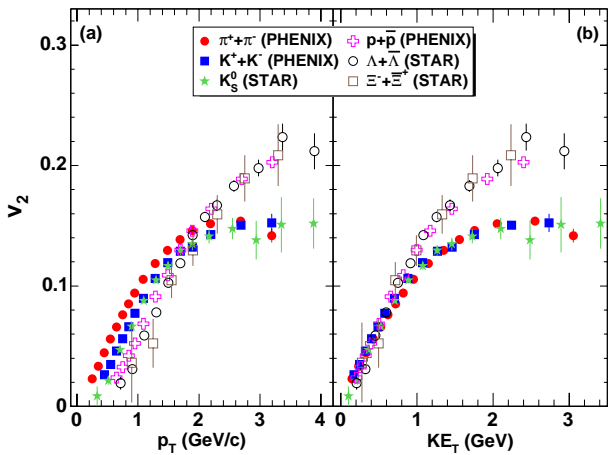}
\caption[]{\label{pid_scaling}
(a) $v_2$ vs $p_T$ and (b) $v_2$ vs $KE_T$ for identified particle 
species obtained in minimum bias Au+Au collisions.  
The STAR data are from Refs.~\cite{Adams:2003am,Adams:2005zg}. 
}
\end{figure}
%

\begin{figure}[tb]
\includegraphics[width=1.0\linewidth]{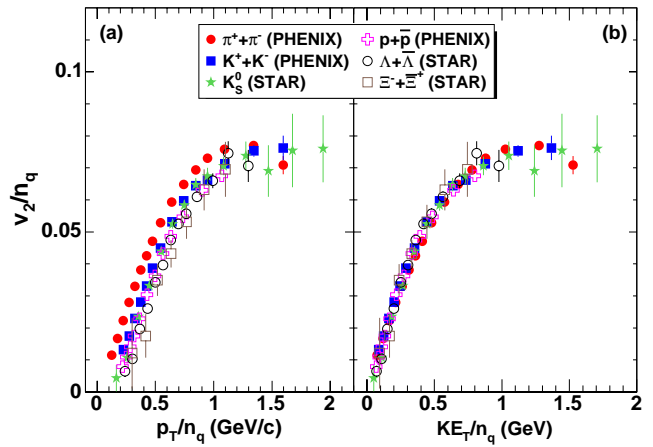}
\caption[]{\label{quark_scaling}
(a) $v_2/n_q$ vs $p_T/n_q$ and (b) $v_2/n_q$ vs $KE_T/n_q$ 
for identified particle species obtained in minimum bias Au+Au collisions. 
The STAR data are from Refs.~\cite{Adams:2003am,Adams:2005zg}. 
}
\end{figure}

Figures \ref{pid_scaling} and  \ref{quark_scaling} show that 
the distinctive features of the $v_2$ for identified 
particles provide another detailed set of scaling tests.
Fig.~\ref{pid_scaling}(a) shows a comparison of the measured differential 
anisotropy $v_2(p_T)$, for several particle species obtained in 
minimum bias Au+Au collisions at $\sqrt{s_{NN}} = 200$~GeV. 
The results 
are in good agreement (better than 3\%) with 
those of our previous measurements \cite{Adler:2003kt}. The values for 
neutral kaons ($K^0_s$), lambdas ($\Lambda$) and the cascades ($\Xi$) show results from the STAR 
collaboration \cite{Adams:2003am,Adams:2005zg}. The STAR $v_2$ values were 
multiplied by the factor 1.1 to account for a small difference 
between the average centralities  
for minimum bias events from the two experiments. 
PHENIX and STAR $v_2(p_T)$ results (for $\pi^{\pm}$, p($\bar{p}$) and $K$) 
for 10\% centrality bins are essentially identical.

The comparison in Fig.~\ref{pid_scaling}(a) shows the well 
known particle identification (PID) ordering of $v_2(p_T)$ at both low and high $p_T$ values.  
At low $p_T$ ($p_T \alt 2$ GeV/$c$), one can see rather clear evidence for mass ordering.
If this aspect of $v_2$ is driven by a hydrodynamic 
pressure gradient, the prediction is that the differential $v_2$ values 
observed for each particle species should scale with $KE_T$. 
The pressure gradient that drives elliptic flow is 
directly linked to the collective kinetic energy of the emitted particles.
For higher values of $p_T$ ($p_T \sim 2-4$ GeV/$c$), Fig.~\ref{pid_scaling}(a) indicates
that mass ordering is broken and $v_2$ is more strongly 
dependent on the quark composition of the particles than on their mass, 
which has been attributed to the dominance of the quark 
coalescence mechanism for $p_T \sim 2-4$ GeV/$c$ \cite{Fries:2003kq,Adler:2003kt,Adams:2003am}. 

	 	 Figure~\ref{pid_scaling}(b) shows the same $v_2$ data presented in Fig.~\ref{pid_scaling}(a)
plotted as a function of $KE_T$. 
Note that $KE_T$ is a robust 
scaling variable because it takes into account relativistic effects, which are 
especially important for the lightest particles. 
In contrast to the PID ordering observed 
in Fig.~\ref{pid_scaling}(a), 
all particle species scale 
to a common set of elliptic flow values for $KE_T \alt 1$~GeV, confirming 
the strong influence of hydrodynamic pressure gradients. 
For $KE_T \agt 1$~GeV, this particle mass scaling (observed for all particle species) gives 
way to a clear splitting into a meson branch (lower $v_2$) and 
a baryon branch (higher $v_2$). Since both of these branches show rather good 
scaling separately, we interpret this as an initial hint for the 
degrees of freedom in the flowing matter at an early stage. 
	
Figure~\ref{quark_scaling} shows the results 
obtained after quark number scaling of the $v_2$ values shown 
in Fig. \ref{pid_scaling}. That is,
$v_2$, $p_T$ and $KE_T$ are divided 
by the number of 
constituent quarks $n_q$ for mesons ($n_q = 2$) and baryons ($n_q = 3$).
Fig. \ref{quark_scaling}(a) indicates rather poor scaling for $p_T/n_q \alt 1$ GeV/$c$ and 
much better scaling for  $p_T/n_q \agt 1.3$ GeV/$c$, albeit with large error bars. 
In contrast, Fig.~\ref{quark_scaling}(b) shows excellent 
scaling over the full range of $KE_T/n_q$ values. We interpret this as an  
indication of the inherent quark-like degrees of freedom in the flowing matter. 
These degrees of freedom are gradually revealed as $KE_T$ 
increases above $\sim 1$~GeV (cf. Fig.~\ref{pid_scaling}(b)) and are apparently 
hidden by the strong hydrodynamic mass scaling, which predominates at low $KE_T$. 
The fact that $v_2/n_q$ shows such good scaling over the entire range of $KE_T/n_q$ 
and does not for $p_T/n_q$, serves to highlight the fact that hydrodynamic 
mass scaling is preserved over the domain of the linear increase 
in $KE_T$. Fig.~\ref{quark_scaling}(b) should serve to distinguish between 
different quark coalescence models.   
%

	In summary, we have presented the results from detailed tests of 
hydrodynamic scaling of azimuthal anisotropy
in Au+Au and Cu+Cu collisions at \sqrtsNN = 200 GeV. For a broad range of 
centralities, eccentricity scaling is observed for charged hadrons for 
both the Cu+Cu and Au+Au systems. For a given eccentricity, $v_2$ is also 
found to be independent of colliding system size.
The observed scaling for identified particles in Au+Au collisions, coupled
with $\epsilon$ scaling, gives strong evidence for hydrodynamic scaling of
$v_2$ over a broad selection of the elliptic flow data. For $KE_T \sim
1-4$~GeV universal hydrodynamic scaling is violated, but baryons and mesons
are found to scale separately. Quark number scaling ($v_2/n_q$ vs.
$KE_T/n_q$) in this domain leads to comprehensive overall scaling of the
data, with substantially better scaling behavior than that found for
$v_2/n_q$ vs. $p_T/n_q$. The scaling with valence quark number may indicate 
a requirement of a minimum number of objects in a localized space that contain 
the prerequisite quantum numbers of the hadron to be formed.  Whether the 
scaling further indicates these degrees of freedom are present at the earliest 
time is in need of more detailed theoretical investigation.
%
%


We thank the staff of the Collider-Accelerator and
Physics Departments at BNL for their vital contributions.
We acknowledge support from
the Department of Energy and NSF (U.S.A.),
MEXT and JSPS (Japan),
CNPq and FAPESP (Brazil),
NSFC (China),
MSMT (Czech Republic),
IN2P3/CNRS, and CEA (France),
BMBF, DAAD, and AvH (Germany),
OTKA (Hungary),
DAE (India),
ISF (Israel),
KRF and KOSEF (Korea),
MES, RAS, and FAAE (Russia),
VR and KAW (Sweden),
U.S. CRDF for the FSU,
US-Hungarian NSF-OTKA-MTA,
and US-Israel BSF.

%

%
%
\end{document}